\documentclass[aps,prb,showpacs,reprint, superscriptaddress]{revtex4-1}

\usepackage{graphicx}
\usepackage{color}
\usepackage{amsmath}
\usepackage{bbm}

\input epsf

\begin{document}

\title{Superconducting dome in doped 2D superconductors with broken inversion symmetry}

\author{P. W\'ojcik}
\email{pawel.wojcik@fis.agh.edu.pl}
\affiliation{AGH University of Science and Technology, Faculty of
Physics and Applied Computer Science, 30-059 Krakow, Poland
Al. Mickiewicza 30, 30-059 Krakow, Poland}

\author{M. P. Nowak}
\affiliation{AGH University of Science and Technology, Academic Centre
for Materials and Nanotechnology, Al. Mickiewicza 30, 30-059 Krakow,
Poland}

\author{M. Zegrodnik}
\affiliation{AGH University of Science and Technology, Academic Centre
for Materials and Nanotechnology, Al. Mickiewicza 30, 30-059 Krakow,
Poland}

\begin{abstract}
We analyze an unconventional inter-subband paired phase in a 2D doped superconductor considering both systems  with the inversion symmetry and with the inversion symmetry broken. We find that for a centro-symmetric system the inter-subband pairing can appear in the high concentration regime when the repulsive Coulomb interaction leads to the nearly degenerate symmetric and antisymmetric state.  We discuss in detail the mutual competition between the intra- and inter-subband paired phase. For systems with broken inversion symmetry, we find that the critical temperature has a characteristic domelike shape as a function of the asymmetry parameter, which is explained as resulting form the inter-subbband pairing. This results is discussed in the context of the domelike shape of $T_c$ in the LaAlO$_3$/SrTiO$_3$ interface.
\end{abstract}

\maketitle
%%%%%%%%%%%%%%%%%%%%%%%%%%%%%%%%%%%%%%%%%%%%%%%%%%%%%%%%%%%%%%%%%%%%%%%%
%introduction
\section{Introduction}
Since two-band superconductivity has been experimentally reported in MgB$_2$\cite{Souma2003,Giubileo2001,Chen2001,Szabo2001}, the multiband superconductors have attracted the growing interest due to 
their unique properties such as vortices structure with fractional flux\cite{Babaev2002} or an enhancement of the upper critical field and the 
critical temperature\cite{Gurevich2003}. The problem of BCS pairing in two bands with different strength of the electron-phonon interaction was resolved already more than 50 years ago by Suhl 
\textit{et al.}\cite{Suhl1959} who reported the appearance of two different superconducting gaps corresponding to the two orbitals. Only at the beginning of this century such phenomenon was confirmed 
experimentally in MgB$_2$ where the two gaps are related respectively to the $\pi$ band formed by the $p_z$ orbitals and $\sigma$ band constituted by a linear combination of the orbitals $p_x$ and 
$p_y$\cite{Szabo2001}. Now, multiband superconductivity is found in many compounds ranging from iron-based high-$T_c$ superconductors\cite{Kuroki2008, Dai2008, Jeglic2010, Singh2008} to a 
two-dimensional electron 
gas at LaAlO$_3$/SrTiO$_3$ interfaces\cite{Joshua2012, Reyren2007, Scheurer2015, Trevisan2018}. In many of them the elucidation of superconducting properties are the subject of the ongoing debate 
with the special attention directed towards the interband pairing and its role in superconductivity in those materials\cite{Dai2008,Tahir1998,Dolgov1987}.

In the last decade, the rapid progress in growth and characterization techniques has pushed the study on multiband superconductivity towards nanostructures. In the nanoscale regime, the Fermi surface 
splits into series of subbands due to the quantum size effect, which naturally leads to multiband superconductivity similar to that observed in MgB$_2$. The multiband character of 
superconductivity in nanostructures was confirmed  by measurements of the thickness dependent oscillations of the critical temperature\cite{Ozer2006,Eom2006,Guo2004} which were explained as resulting 
from the van Hove sigularities occurring each time when the bottom of the subband passes through the Fermi level (Lifshitz transition)\cite{Shanenko2008,Shanenko2007,Wojcik2014}. Interestingly, 
recent experiments show that the electron-phonon coupling constant varies from one subband to another and might be different even for two neighboring subbands\cite{Matetskiy2015}. A determination 
of the electron-phonon coupling in individual subbands and between them in nanostructures is still an open issue, from both theoretical and experimental point of view.
In particular, when the energy between electronic states (subbands) becomes smaller than the Debay window\cite{Shanenko2015} the unconventional inter-subband pairing could appear.

In this paper we present that the latter condition can be satisfied in 2D doped superconductors (nanofilm) in the high concentration regime when the repulsive Coulomb interaction leads to the 
formation of nearly degenerate electronic states that correspond to symmetric and antisymmetric wave-functions in the growth direction. We consider both the system with the inversion 
symmetry and the situation when the inversion symmetry is broken by the external potential. Interestingly, for the 
latter case we found that the critical temperature has a characteristic domelike shape which is explained as resulting form the inter-subbband pairing.

The structure of the paper is as follows. In section \ref{sec:2} we introduce the theoretical model based on the coupled Poisson and BCS equations. In section \ref{sec:3} we present our results for 
both symmetric and asymmetric quantum confinement. Discussion with reference to superconductivity at LaAlO$_3$/SrTiO$_3$ interfaces is included in section \ref{dis:sto} while conclusions are provided 
in section \ref{sec:4}.

\section{Theoretical model}
\label{sec:2}
In this work, we solve the BCS model for a thin doped 2D superconductor with the inclusion of the electron-electron interaction treated in the mean-field approximation. The long-range Coulomb and the 
superconducting contact interaction can be combined in one formalism via Hubbard-Stratonovich transformation\cite{Otterlo1995,Ambegaokar1982}. Considering only the classical saddle point approximation 
in the static limit, the variation with respect to the electrostatic potential $\phi$ and the order parameter $\Delta$ (for more details see Ref.~\onlinecite{Virtanen2019}) gives the coupled Poisson 
equation 
\begin{equation}
 \nabla ^2 \phi(\mathbf{r})=\frac{e n_e(\mathbf{r})}{\epsilon _0 \epsilon},
 \label{eq:poisson}
 \end{equation}
 and the BCS Hamiltonian in the form
 \begin{eqnarray}
\hat{\mathcal{H}}_{BCS}&=&\sum _{\sigma} \int d^3 r
\hat{\Psi}^{\dagger} (\mathbf{r},\sigma) \hat{H}_e ^{\sigma}
\hat{\Psi}(\mathbf{r},\sigma) \nonumber \\ 
&+& \int d^3 r \left [ \Delta
(\mathbf{r})\hat{\Psi}^{\dagger}(\mathbf{r},\uparrow)
\hat{\Psi}^{\dagger}(\mathbf{r},\downarrow) +H.c. \right ] \nonumber \\
&+&\int d^3r \frac{|\Delta(\mathbf{r})|^2}{g},
\label{eq:ham}
\end{eqnarray}
where $\sigma$ corresponds to the spin state $(\uparrow, \downarrow)$, $g$ is the electron-electron coupling constant and $\hat{H}_e^{\sigma}$ is the single-electron Hamiltonian given by
\begin{equation}
 \hat{H}_e^{\sigma} = -\frac{\hbar^2}{2m}\nabla ^2 - e\phi(\mathbf{r}) + U(\mathbf{r})-\mu,
\end{equation}
where $\mu$ is the Fermi energy and $U(\mathbf{r})$ is the external potential.

The superconducting gap parameter $\Delta(\mathbf{r})$ in real space is defined as
\begin{equation}
 \Delta(\mathbf{r})=-g \left < \hat{\Psi} (\mathbf{r},\downarrow)
\hat{\Psi} (\mathbf{r},\uparrow)  \right >.
\label{eq:gap_def}
\end{equation}
with the field operators 
\begin{equation}
 \hat{\Psi}(\mathbf{r},\sigma)=\sum_{n,\mathbf{k}} \psi_{n,\mathbf{k}}(\mathbf{r})\:\hat{c}_{n, \mathbf{k}, \sigma},
\label{eq:field_op}
\end{equation}
where $\hat{c}_{n,\mathbf{k}, \sigma} (\hat{c}^{\dagger}_{n, \mathbf{k}, \sigma})$ is
the anihilation (creation) operator for an electron with spin $\sigma$ in the subband $n$ characterized by
the wave vector $\mathbf{k}$ and $\psi_{n, \mathbf{k}}(\mathbf{r})$ are the single-electron eigenfunctions  of the Hamiltonian $\hat{H}^{\sigma}_e$.

The superconductivity in the considered system is determined by either the intra-subband and the inter-subband pairing. The multiband BCS Hamiltonian takes the form 
\begin{equation}
\begin{split}
\hat{\mathcal{H}}_{BCS}&=
\sum_{\mathbf{k}}\mathbf{\hat{f}}^{\dagger}_{\mathbf{k}} H_{\mathbf{k}}\mathbf{\hat{f}}_{\mathbf{k}}+\sum_{n,\mathbf{k}} \xi_{n, -\mathbf{k},\downarrow}  + \sum_{n,m} \frac{|\Delta_{n,m}|^2}{g},  
\end{split}
\label{eq:Ham_matrix_1}
\end{equation}
where $\mathbf{\hat{f}}^{\dagger}_{\mathbf{k},\mathbf{Q}}=(\hat{c}^{\dagger}_{1,\mathbf{k},\uparrow}, \hat{c}_{1,-\mathbf{k},\downarrow}, \dots, \hat{c}^{\dagger}_{N,\mathbf{k},\uparrow}, \hat{c}_{N,-\mathbf{k},\downarrow})$ is the composite vector operator ($N$ determines the number of subbands which participate in superconductivity)  and
\begin{equation}
H_{\mathbf{k}}=\left(\begin{array}{ccccc}
\xi_{1,\mathbf{k},\uparrow} & \Gamma_{1,1} & \dots & 0 & \Gamma_{1,N}\\
\Gamma_{1,1} & -\xi_{1, -\mathbf{k}, \downarrow} & \dots & \Gamma_{N,1} & 0 \\
\vdots & \vdots & \ddots & \vdots & \vdots \\
 0 & \Gamma_{N,1} & \dots & \xi_{N,\mathbf{k},\uparrow} & \Gamma_{N,N} \\
\Gamma_{1,N} & 0 & \dots  & \Gamma_{N,N}& -\xi_{N, -\mathbf{k}, \downarrow}
\end{array} \right).
\label{eq:matrix_H}
\end{equation}
The intra- and inter-subband superconducting gap parameters, $\Gamma_{n,m}$, are expressed by 
\begin{equation}
 \Gamma_{n,m} = -g \sum _{n',m'} V_{n,m}^{n',m'} \Delta _{n',m'},
 \label{eq:gamma}
\end{equation}
where the interaction matrix elements 
\begin{equation}
 V_{n,m}^{n',m'}=\int d^3r \psi^*_{n,\mathbf{k}}(\mathbf{r}) \psi^*_{m,\mathbf{k}}(\mathbf{r}) \psi_{n',\mathbf{k}}(\mathbf{r}) \psi_{m',\mathbf{k}}(\mathbf{r})
\label{eq_inter}
\end{equation}
and 
\begin{equation}
\Delta _{n,m} = \sum_k{}^{'}  \langle \hat{c}_{n, \mathbf{-\mathbf{k}}, \downarrow} \hat{c}_{m, \mathbf{k}, \uparrow } \rangle.
\label{eq:delta}
\end{equation}
The summation in Eq. (\ref{eq:delta}) is carried out only if both the single electron states $\xi_{n,\mathbf{k}}$ and $\xi_{m,\mathbf{k}}$ are located inside the cutoff window $[\mu-E_{cut}, \mu + E_{cut}]$, where $E_{cut}$ is the cutoff energy.

The single-electron eigenfunctions $\psi_{n,\mathbf{k}}(\mathbf{r})$ and energy $\xi_{n,\mathbf{k}}$ as well as the electrostatic potential $\phi(\mathbf{r})$ are calculated through the solution of the Schr{\"o}dinger-Poisson problem in the constant electron concentration regime.
 %Eq.~(\ref{eq:poisson}), with
%\begin{equation}
%n_e(\mathbf{r})=\sum _{n,\mathbf{k}} 2 | \psi _{n,\mathbf{k}} (\mathbf{r}) |^2 f(\xi_{n,%\mathbf{k}}),
%\label{eq:ne}
%\end{equation}
%where $f(\xi_{n,\mathbf{k}})$ is the Fermi-Dirac function. The use of the Fermi function instead of the quasiparticle distribution in Eq.~(\ref{eq:ne}) is a widely used approximation\cite{Shanenko2008}, 
%weak point justified when $\mu>>E_{cut}$.

The system under consideration has a form of the free-standing nanofilm with the thickness $d$. We assume that the system is infinite in the $x-y$ plane while in the $z$-direction the hard wall boundary conditions are adopted, i.e. $\psi_{n,\mathbf{k}}(0)=\psi_{n,\mathbf{k}}(d)=0$. In the $x-y$ plane we use the parabolic band approximation for which $V_{n,m}^{n',m'}$ do not depend on the $\mathbf{k}$ vector. The calculations are carried out in the following manner: for a given electron concentration $n_e$ and the potential $U(\mathbf{r})$ we solve the Schr{\"o}dinger-Poisson problem to determine $\psi_{n,\mathbf{k}}(\mathbf{r})$, $\xi_{n,\mathbf{k}}$ and $\phi(\mathbf{r})$. These quantities are then used in the BCS model. The numerical diagonalization of (\ref{eq:matrix_H}) leads to the quasiparticle energies which are exploited to calculate the paring energies $\Gamma _{n,m}$ by solving the set of self-consistent equations (\ref{eq:matrix_H})-(\ref{eq:gamma}).

The calculations were carried out for the following parameters: $m=1$, $\epsilon=1$, $gN(0)=0.18$ and $E_{cut}=32.31$~meV which gives $\Gamma _{bulk}=0.25$~meV.

\section{Results}
\label{sec:3}
\begin{figure*}[ht]
\begin{center}
\includegraphics[scale=0.4]{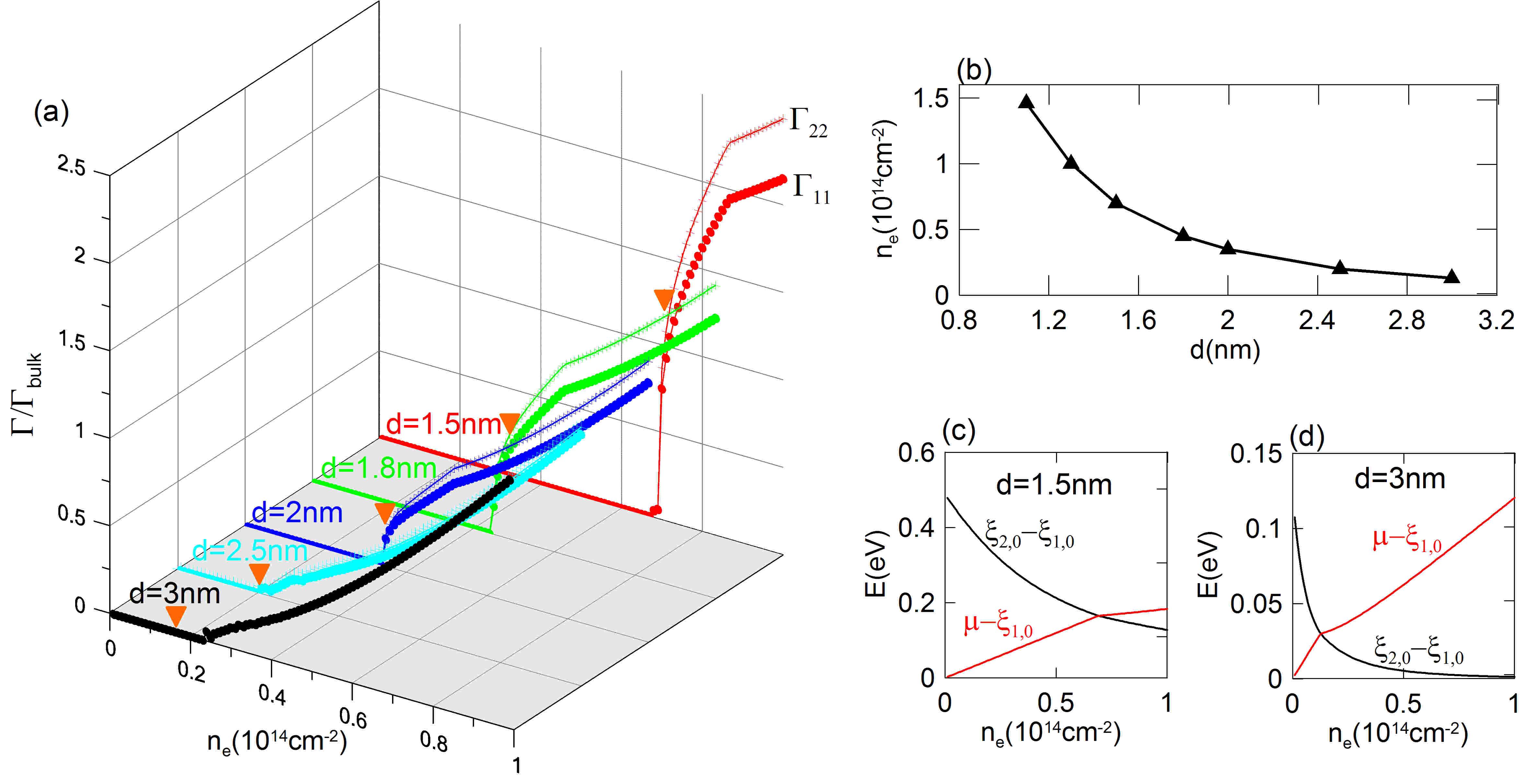}
\caption{(a) The intra-subband pairing energy $\Gamma_{1,1}$ (thick lines) and $\Gamma_{2,2}$ (thin lines) as a function of the electron concentration $n_e$ for different nanofilm thicknesses $d$. (b) The critical electron concentration $n_e^c$ as a function of the nanofilm thickness $d$. (c,d) The bottom of the second subband $\xi_{2,0}$ and the chemical potential $\mu$ calculated with respect to the bottom of the first subband $\xi_{1,0}$ \textit{vs.} the electron concentration $n_e$ for $d=1.5$~nm and $d=3$~nm.} 
\label{fig1}
\end{center}
\end{figure*}

In the first part of our analysis, we consider a nanofilm with the inversion symmetry, putting $U(\mathbf{r})=0$.  
In Fig.~\ref{fig1}(a) we show the intra-subband pairing energy $\Gamma _{i,i}$ as a function of the electron concentration $n_e$ for different nanofilm thicknesses $d$. The increase of the electron concentration leads to occupation of subsequent subbands whereas for the electron density and nanofilm thickness range presented in Fig.~\ref{fig1}(a), maximally two lowest subbands can be filled when increasing $n_e$. For each thicknesses $d$, the Lifshitz transition, appearing when the second subband passes the Fermi level (marked in Fig.~\ref{fig1}(a) by the orange triangle), is accompanied with the enhancement of $\Gamma _{11}$ and $\Gamma _{22}$ in accordance with the BCS theory - a rapid increase of the electronic density of states (DOS) leads to the enhancement of the superconducting energy gap. 
The critical electron concentration $n_e^c$, at which the Lifshitz transition occurs, moves toward lower values with increasing $d$, as presented in Fig.~\ref{fig1}(b). This critical density $n_e^c$ was determined as a crossing point between the energy of the second subband $\xi_{2,0}$ and the chemical potential, both calculated with respect to the energy of the first subband $\xi_{1,0}$ - see Fig.\ref{fig1}(c,d). Note that for both nanofilm thicknesses presented in Fig.~\ref{fig1}(c,d), at $n_e^c$ we observe a change of the slope in the $\mu(n_e)$ dependence characteristic for the Lifshitz transition. 
\begin{figure}[ht]
\begin{center}
\includegraphics[scale=0.45]{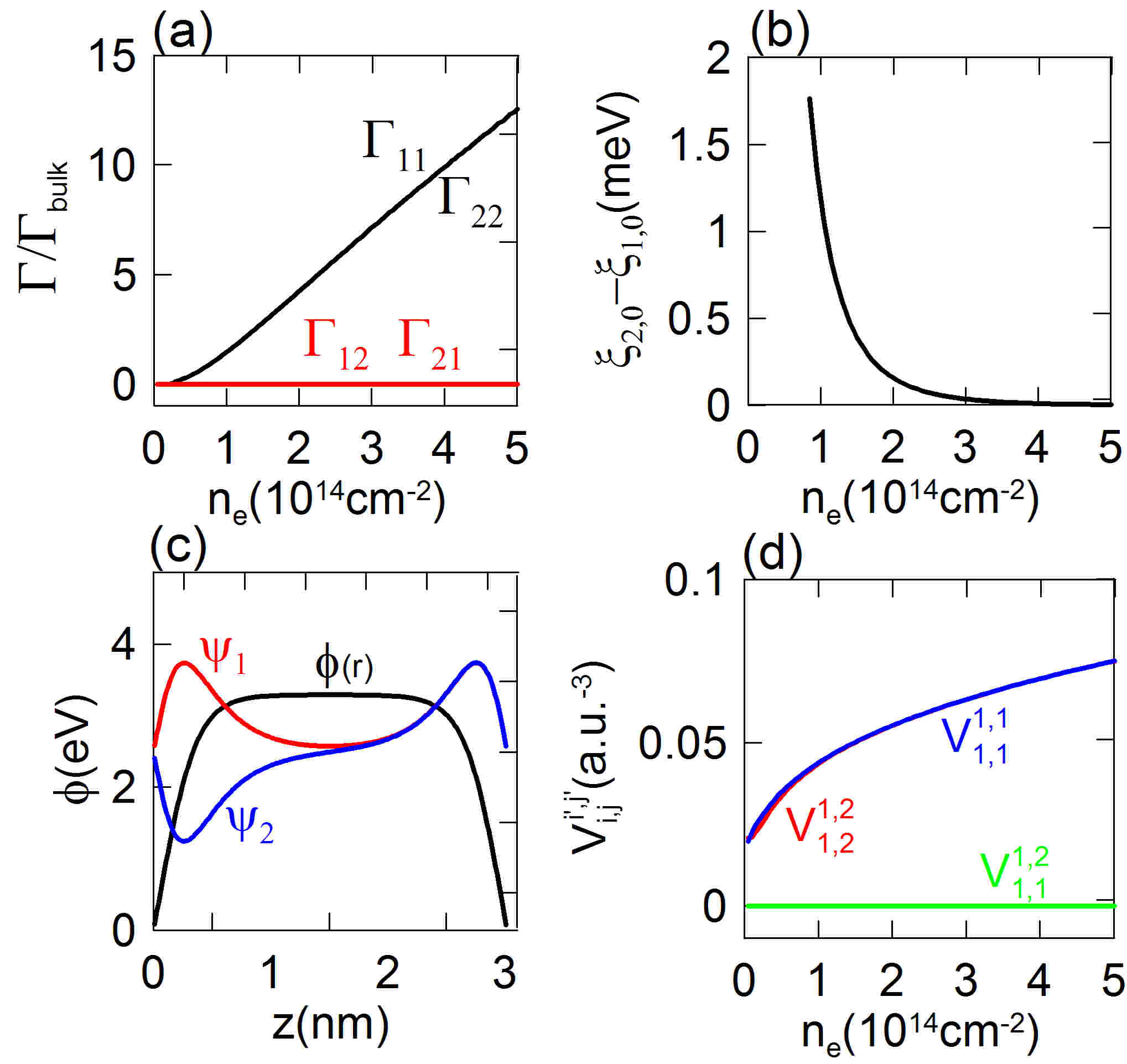}
\caption{(a) The intra- ($\Gamma_{1,1(2,2)}$) and inter-subband ($\Gamma_{1,2(2,1)}$) pairing energy as a function of the electron concentration $n_e$. (b) The difference between the energy of the second $\xi_{2,0}$ and the first $\xi_{1,0}$ subband as a function of the electron concentration $n_e$. (c) The Hartree potential $\phi(z)$ together with wave functions of the two lowest electronic states $\psi _1$ and $\psi _2$ for $n_e=10^{14}$~cm$^{-2}$. (d) Non-zero interaction matrix elements $V_{i,j}^{i',j'}$ \textit{vs.}  $n_e$. Results for $d=3$~nm.} 
\label{fig2}
\end{center}
\end{figure}

In the considered system both the intra- and inter-subband pairing is determined by the electron-electron attraction strength which effectively depends on the coupling constant $g$ and the interaction matrix elements $V_{n,m}^{n',m'}$, Eq~(\ref{eq:gamma}). Note that to generate the pairing correlations, the single electronic states which form the Cooper pairs must be located within the cutoff energy window. This limits the number of subbands that can pair only to those between which the energy difference does not exceed $E_{cut}$. However, this condition is not sufficient for the case of the inter-subband paring as the weak electron-electron coupling cannot generate the superconducting correlations between the two subbands with a significant Fermi wave vector mismatch. For this reason the inter-subband pairing can be observed only between the bands which are energetically close to each other. Below, we show that this condition can be fulfilled in the considered 2D doped superconductors as an effect of the electron-electron Coulomb interaction.

The behavior of the electronic states in the considered quasi-2D nanofilm results from a complex interplay between the quantum confinement and the self-consistent field due to the electron-electron interaction. In the low electron concentration regime, the electron-electron interaction is negligible, quantum confinement effect dominates electronic energies and the Hartee potential $\phi(z)$ is nearly flat (not shown here). As a results, the wave function $\psi _1$ and $\psi _2$ are localized in the centre of the quantum well. This, in turn, leads to the intra-subband interaction matrix elements $V_{n,n}^{n,n}$ larger than the inter-subband correspondents $V_{n,m}^{n,m}$, what makes the intra-subband pairing more preferable. The inter-subband pairing is additionally weakened by the considerable energy separation between the subbands ($\approx 3\hbar^2\pi^2 / 2md^2$) which results in the wave vector mismatch between the states $|n,k \rangle$ and $|m,-k\rangle$ at the Fermi level.

In the high concentration regime the electron-electron interaction dominates and the total energy is minimised by reducing repulsive Coulomb energy. As shown in Fig.~\ref{fig2}(c), the Hartree potential takes the from of a symmetric potential barrier localized in the middle of the quantum well. As a result, electrons are pushed apart and accumulate near the surfaces creating nearly degenerate symmetric and antisymmetric states -- see Fig.~\ref{fig2}(c). The difference in energy between both the electronic states $\psi _1$ and $\psi _2$ as a function of the electron concentration $n_e$ is presented in Fig.~\ref{fig2}(b). Note that the dominant electron-electron interaction makes the inter-subband pairing preferable also by the change of the interaction matrix elements. In general, due to the symmetry of the single-electron eigenfunctions $V_{n,n}^{n,m}=V_{n,n}^{m,n}=V_{n,m}^{m,m}=V_{n,m}^{n,n}=0$ and $V_{n,n}^{m,m}=V_{m,m}^{n,n}=V_{n,m}^{n,m}=V_{n,m}^{m,n}$. This, in turn, reduces the interaction matrix to three non-zero elements: two intra-subband $V_{1,1}^{1,1}$ and $V_{2,2}^{2,2}$ and the inter-subband $V_{1,2}^{1,2}=V_{2,1}^{2,1}=V_{1,1}^{2,2}=V_{2,2}^{1,1}$, which, as presented in Fig.~\ref{fig2}(d), become equal for the nearly degenerate symmetric and antisymmetric states. 
\begin{figure}[ht]
\begin{center}
\includegraphics[scale=0.35]{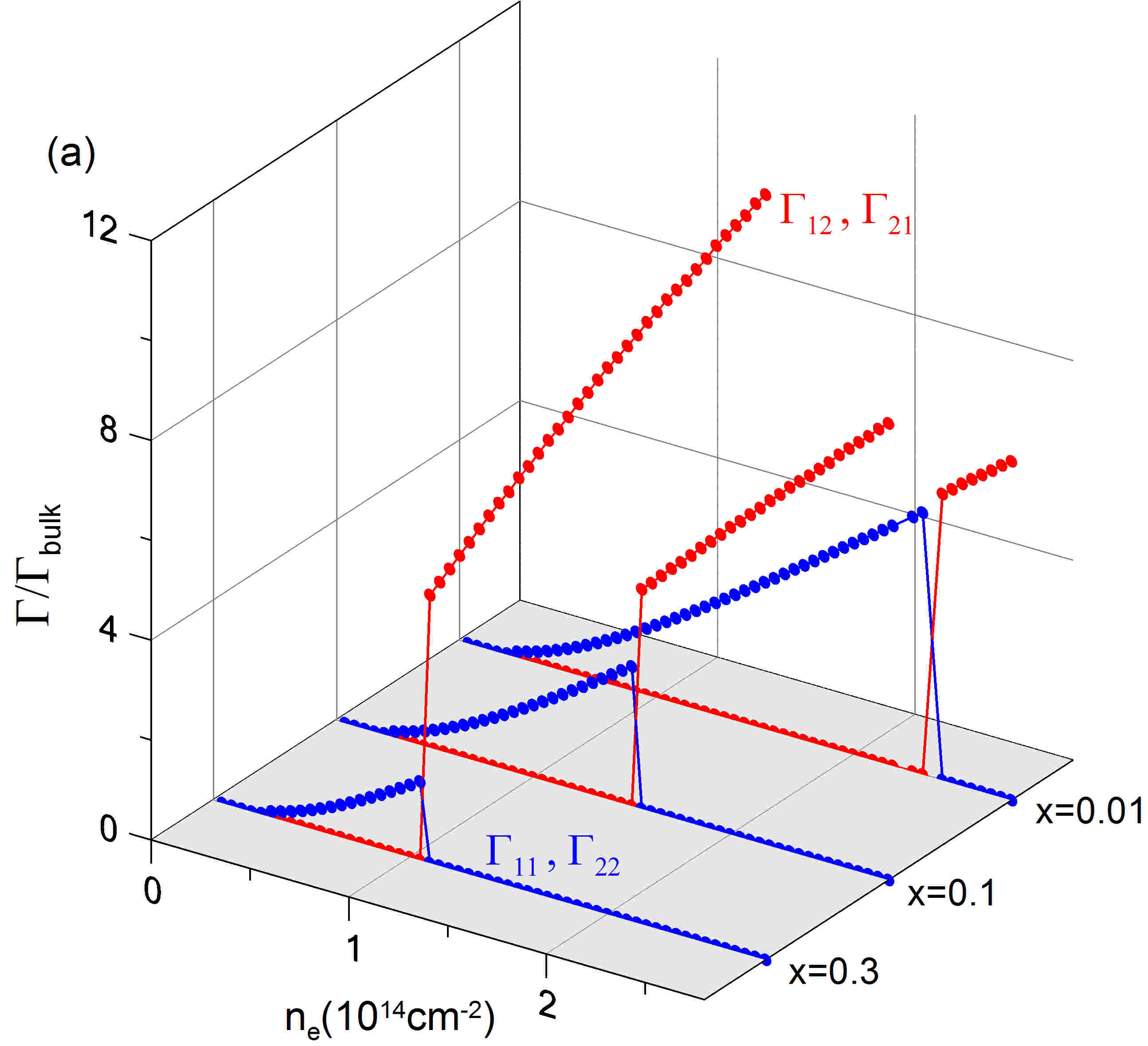}
\caption{(a) The intra- ($\Gamma_{1,1(2,2)}$) and inter-subband ($\Gamma_{1,2(2,1)}$) pairing energy as a function of the electron concentration $n_e$ for different asymmetry factors $x$.} 
\label{fig3}
\end{center}
\end{figure}

Fig.~\ref{fig2}(a) presents the intra- and inter-subband pairing energy as a function of the electron concentration calculated up to the value $5\times 10^{14}$~cm$^{-2}$. Although the two subbands are nearly degenerate and the interaction matrix elements for the intra- and inter-subband coupling are equal, the energy of the inter-subband coupling $\Gamma_{1,2(2,1)}$ is zero for the whole considered electron concentration regime. This unexpected behavior can be explained as an effect of the inversion symmetry of system, which introduces a strong detrimental influence of the intra-subband pairing on the inter-subband pair formation. This fact can be understood based on Eq.~(\ref{eq:gamma}) whose explicit form for the system with the inversion symmetry is given by
\begin{equation}
\begin{split}
\Gamma_{11(22)}&=-g(V\Delta_{11(22)}+V^\prime\Delta_{22(11)}), \\
\Gamma_{12(21)}&=-g(V^\prime \Delta_{12(21)}+V\Delta_{21(12)}),
\label{eq:gamma2}
\end{split}
\end{equation}
where we denote $V_{1,1}^{1,1} = V_{2,2}^{2,2} = V$ and $V_{1,2}^{1,2}=V_{2,1}^{2,1}=V_{1,1}^{2,2}=V_{2,2}^{1,1}=V^\prime$.
The terms with $V^\prime \Delta_{12(21)}$ and $V^\prime\Delta_{21(12)}$ correspond to the inter-subband pairing, while $V^\prime\Delta_{22(11)}$ refers to the inter-subband pair hopping. The latter is operative only when the intra-subband pairs are created ($\Delta_{11}\neq 0$ and $\Delta_{22}\neq 0$). In such a case and when the symmetry of the Cooper pairs tunneling rate between the bands is lifted due to the energy separation, the $V^\prime \Delta_{22(11)}$ term enhances the disproportion between the electron concentrations in the two subbands~\cite{Moreo2009}. This in turn strongly suppresses the inter-subband pairing and, more importantly, excludes the coexistence of the intra- and inter-subband phase. Nevertheless, in the high concentration regime $V=V^\prime$ [Fig.~\ref{fig2}(d)], the free energy of the intra-subband superconducting state is slightly lower due to the slight difference in energy between the subbands.   
\begin{figure}[ht]
\begin{center}
\includegraphics[scale=0.45]{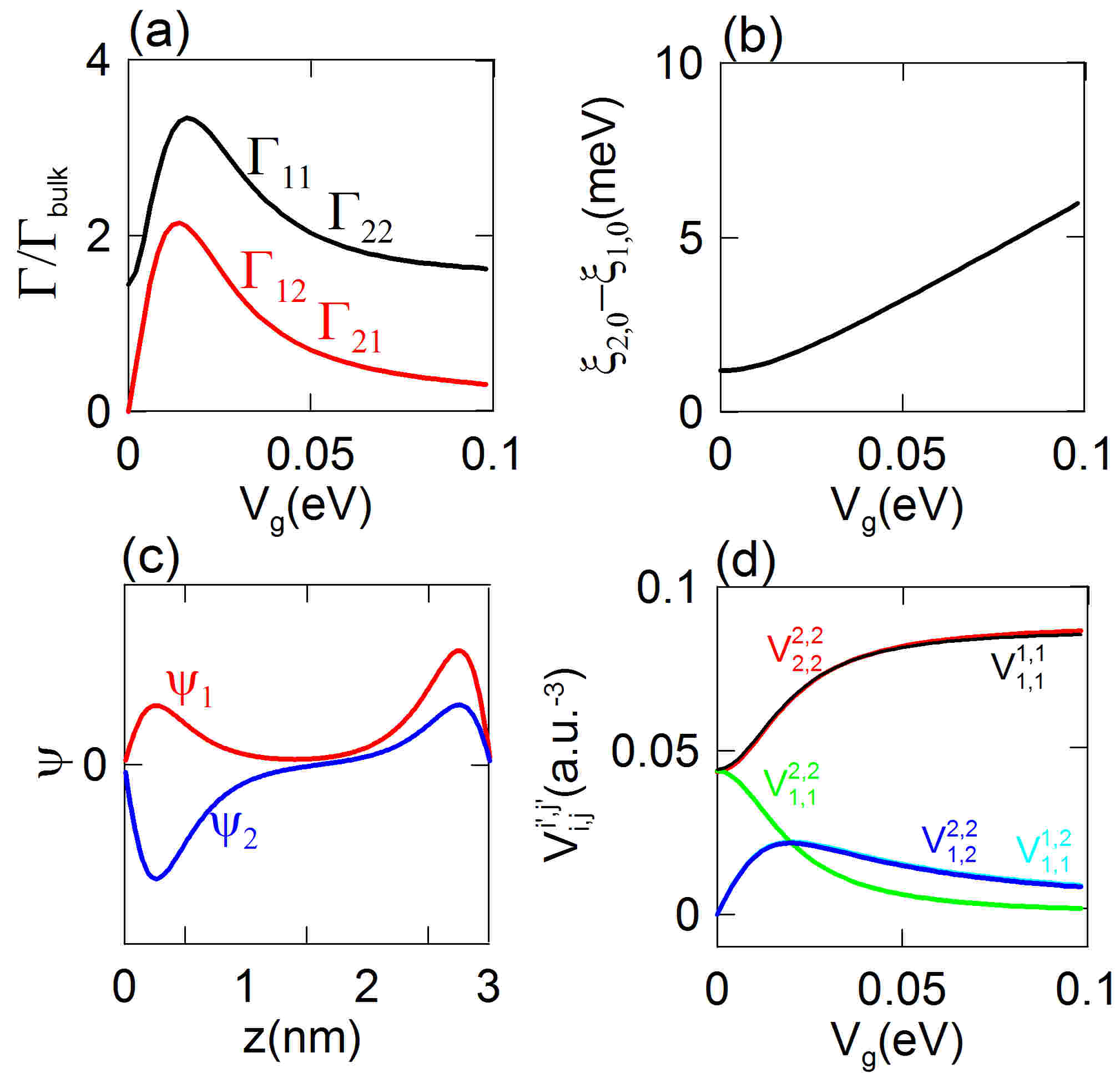}
\caption{(a) The intra- ($\Gamma_{1,1(2,2)}$) and inter-subband ($\Gamma_{1,2(2,1)}$) pairing energy as a function of the parameter $V_g$. (b) The difference between the energy of the second $\xi_{2,0}$ and the first $\xi_{1,0}$ subband \textit{vs.} $V_g$. (c) Wave functions of the two lowest electronic states $\psi _1$ and $\psi _2$. (d) Interaction matrix elements $V_{i,j}^{i',j'}$ as a function of $V_g$. Results for $d=3$~nm and $n_e=10^{14}$~cm$^{-2}$.} 
\label{fig4}
\end{center}
\end{figure}

The relative energy between the phases can be easily altered in favor of the inter-subband phase by introducing a slight asymmetry between $V$ and $V^\prime$. Fig.~\ref{fig3} presents the phase diagrams for different asymmetry factor defined as $x = V^\prime/V-1$. Note that even an extremely small asymmetry at the level of $x=0.01$ leads to the phase transition at some electron concentration which decreases with increasing $x$. Regardless of $x$ the intra- and inter-subband phases do not coexist and the appearance of one phase results in vanishing of the other.

Now, let us analyze the case when the inversion symmetry is broken by introducing the external potential in the simple form $U(z)=-V_gz/d$, where $d$ is the nanofilm thickness and $V_g$ is the parameter which may correspond to the gate voltage. Such form of the $U(z)$ deforms the potential confinement leading to a triangle-shaped quantum well. For the non-centrosymmetric system the single electron wave functions do not have well defined parity [Fig.~\ref{fig4}(c)] and the interaction matrix elements with the odd number of the same index $V_{n,n}^{n,m}, V_{n,n}^{m,n}, V_{m,m}^{m,m}, V_{n,m}^{n,n}$ are no longer equal to zero. Then, the pairing symmetry, which leads to the strong competition between the intra- and inter-subband phase, is broken and both phases can coexist. Fig.~\ref{fig4}(a) displays the intra- and inter-subband pairing energy as a function of the $V_g$ parameter for high electron concentration, $n_e=10^{14}$~cm$^{-2}$, for which we have established that the inter-subband pairing could appear. Indeed, in Fig. \ref{fig4}(a) we see that, for the nonzero $V_g$ both the intra- and inter-subband phases coexist with a prominent domelike shape of the inter-subband phase contribution. As presented in Fig.~\ref{fig4}(d) the behavior of $\Gamma_{1,2(2,1)}(V_g)$ is related to the non-centrosymmetric matrix elements $V_{2,1}^{2,2}$ and $V_{1,1}^{1,2}$ whose dependence on $V_g$ have characteristic domelike shape. Note that the reorganization of the energy levels induced by $V_g$ is not able to generate the characteristic shape of $\Gamma_{1,2(2,1)}(V_g)$ as the energy difference between the states increases with $V_g$ being detrimental to the inter-subband pairing [Fig.~\ref{fig4}(b)]. This additional effect of the asymmetry may be responsible for the decrease of the $\Gamma_{1,2(2,1)}(V_g)$ slope just above the maximum. 
\begin{figure}[ht]
\begin{center}
\includegraphics[scale=0.35]{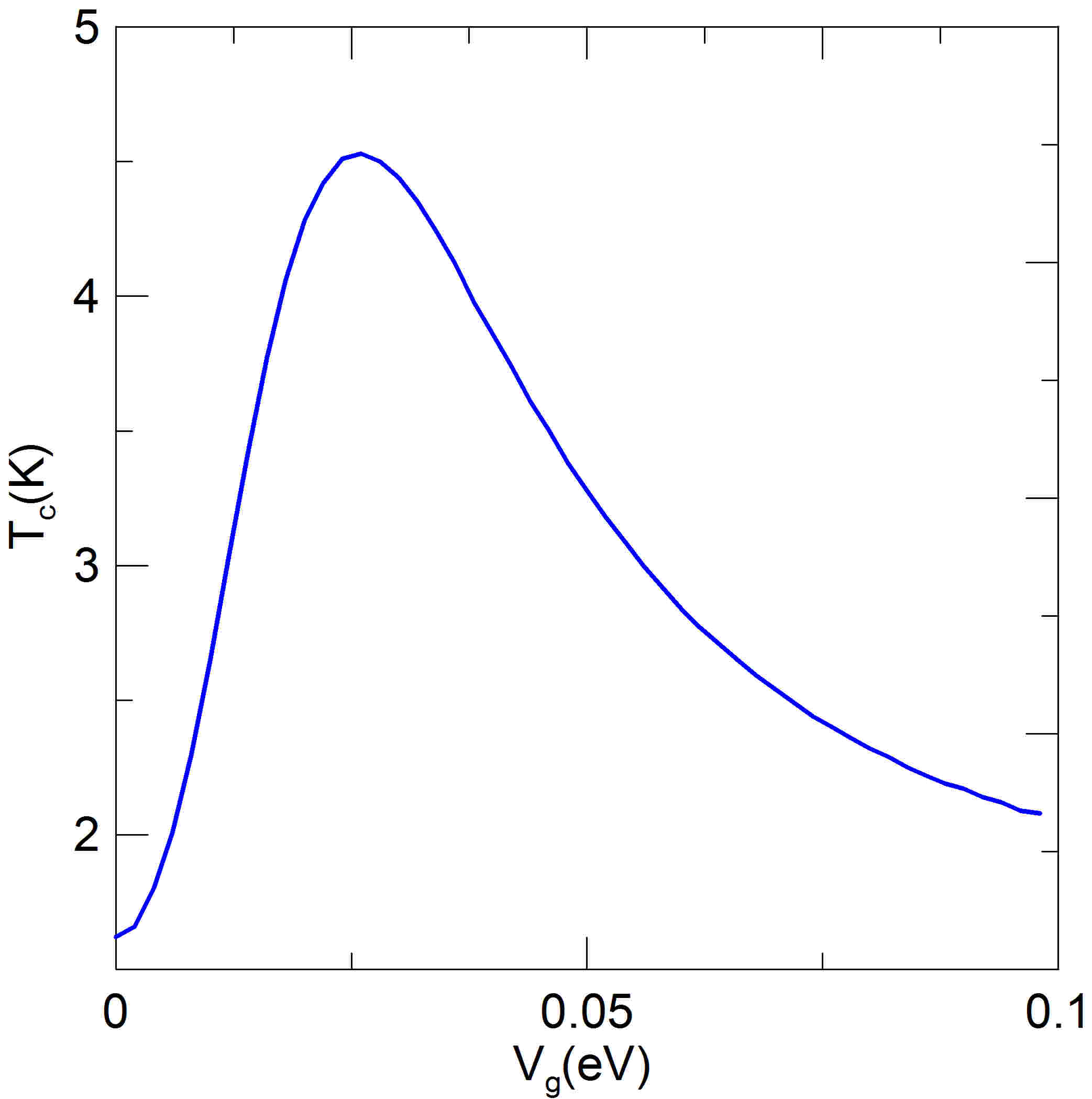}
\caption{Critical temperature $T_c$ as a function of $V_g$ with characteristic domelike shape. Inset: Experimental data for the LaAlO$_3$/SrTiO$_3$ interface, from Ref.~\onlinecite{joshua2012universal}.} 
\label{fig5}
\end{center}
\end{figure}

Finally, we calculate the critical temperature with respect to $V_g$ and present the result in Fig.~\ref{fig5}. We observe that the critical temperature $T_c$ exhibits characteristic domelike shape which, as discussed above, results from the inter-subband pairing induced in a system with broken inversion symmetry.  

\section{Discussion in the context of the LAO/STO interface }
\label{dis:sto}

As we have shown in the previous section in a 2D doped superconductor the critical temperature undergoes domelike dependence on $V_g$. Notably, a similar domelike shape of the critical temperature as a function gate voltage is found for the LaAlO$_3$/SrTiO$_3$ interface - see Fig.~4 in Ref.~\onlinecite{Joshua2012}. For this material the maximum $T_c$ is about $100-200$~mK and appears close to the Lifshitz transition \cite{Reyren1196,Rout2017,joshua2012universal,maniv2015strong,Biscaras,biscaras2010two,caviglia2008electric}. The theoretical explanation of this unique behavior is still an open issue with several theoretical proposals\cite{Monteiro2019,Trevisan2018}. Specifically, the experimental observation contradicts the predictions of the standard BCS model which states that the critical temperature should monotonically increase above the Lifshitz transition due to a sudden increase of electronic states available for the pairing\cite{Shanenko2008,Wojcik2014,Wojcik2015}.

It should be noted that the origin of the pairing mechanism and the effect of electronic correlations on superconductivity in LaAlO$_3$/SrTiO$_3$ is currently widely debated (Ref.~\onlinecite{Trevisan2018} and the references therein). Here we provide an alternative route for explanation of the measured $T_c$ in a model that does not account electron correlations. Already it has been demonstrated that some properties of the LaAlO$_3$/SrTiO$_3$ interfaces can be captured in theoretical analysis that neglect the Coulomb repulsion \cite{Trevisan2018,Valentinis2017,Diez2015} or that apply the mean-field approach, which neglects most of the effects resulting from the inter-electronic correlations\cite{maniv2015strong}.

Even though our model is not directly related to the LaAlO$_3$/SrTiO$_3$ interface, some important similarities between both situations are apparent and allow  to suggest that the inter-subband pairing may have a significant influence on the domelike behavior of $T_c$ at LaAlO$_3$/SrTiO$_3$ interface.
According to our analysis one can formulate two conditions which have to be met in order to observe the characteristic domelike shape driven by the 
inter-subband pairing: (i) the lack of the inversion symmetry and (ii) the energetic proximity of the bands between which the pairing appears. Both requirements are met in the LaAlO$_3$/SrTiO$_3$ interface: (i) the calculations for the conduction band at LaAlO$_3$/SrTiO$_3$ interface based on the Poisson equation demonstrate that the confinement in the $z$-direction at the interface has a shape of the triangular quantum well\cite{Biscaras,Smink,Yin}, which leads to the lack of the inversion symmetry; (ii) the tight-binding model  for the LaAlO$_3$/SrTiO$_3$ structure reveals the presence of three bands two of which ($d_{XY}/d_{YZ}$ and $d_{YZ}/d_{XZ}$) have relatively similar Fermi surfaces just above the Lifshitz transition what makes the inter-subband pairing between them possible\cite{joshua2012universal}. Nevertheless, this proposal still needs a detailed theoretical analysis in a more realistic model with all three $d_{XY}$, $d_{XZ}$, $d_{YZ}$ orbitals, and the influence of the spin-orbit coupling, included.

Finally, the inter-subband pairing origin of the superconducting dome in the LaAlO$_3$/SrTiO$_3$ interface can be verified experimentally from the temperature-dependence of the upper-critical field measured for different gate voltages. As recently shown\cite{Ayino} this dependence displays a positive curvature which is a hallmark of multi-band superconductivity, meaning that the contribution from the inter-subband pairing can be extracted from the appropriate fitting procedure. If indeed the proposed process stands behind the observed $T_c$ dependence on the gate voltage one should observe a maximal value of the inter-subband coupling constant just after the Lifshitz transition. 

\section{Conclusions}
\label{sec:4}
 In the present paper, we have analyzed the inter-subband pairing in a 2D doped superconductor. In order to determine the principal features of the paired state we have used the BCS model where the 
Hatree potential is included in the mean-field approximation by solving the Schr\"odinger-Poisson problem. In the considered model, the strength of the inter-subband paring depends not only on 
the coupling constant $g$ but also on the interaction matrix elements $V_{i,j}^{i',j'}$ which physically determine the overlap between the bands from which the electrons forming the Cooper pairs come from. In order to 
induce the inter-subband paired phase both electronic states should be included in the energy cutoff window while the pairing strength should be sufficient to overcome detrimental influence of the 
Fermi wave vector mismatch. The latter appears as a result of the energy separation between the subbands. This shows that the subbands which are energetically close to each other are favored when it comes to inter-subband pairing. As we presented, the latter condition 
can be satisfied in the high concentration regime when the repulsive Coulomb interaction leads to the nearly degenerate symmetric and antisymmetric state. Even in this regime the inter-subband phase 
does not appear spontaneously in the centrosymmetric system which results from the mutual competition between the intra- and inter-subband phase. Switching between 
those two mutually exclusive phases requires slight asymmetry in the coupling constants. 

Finally, we have shown that for systems with the lack of inversion symmetry, the critical temperature has a characteristic domelike shape as a function of the parameter $V_g$, which defines the shape of the external potential in the $z$-direction. Such behavior has been 
explained as resulting form the inter-subbband pairing which appears due to the nonzero values of the matrix elements $V_{i,i}^{i,j}$ for the non-centrosymmetric systems. Some similarities between our model and the situation which takes place at the LaAlO$_3$/SrTiO$_3$ interface suggest that the the inter-band pairing may play a significant role in the appearance of the domelike shape of $T_c$ as a function of gate voltage. We also propose how this concept can we verified experimentally.

\section{Acknowledgement}
This work was supported by National Science Centre, Poland (NCN) according to decision 2017/26/D/ST3/00109 and in part by PL-Grid Infrastructure.                                                       

%\bibliography{refs.bib}

\end{document}